\documentclass[12pt]{iopart}

\usepackage{graphicx}
\begin{document}

\title{Natural limit on the $\gamma$/hadron separation for a stand alone air Cherenkov telescope}

\author{Dorota Sobczy\'{n}ska}

\address{University of {\L }\'{o}d\'{z}, Experimental Physics 
Department, Pomorska 149/153, 90-236 {\L }\'{o}d\'{z},Poland}
\ead{ds@kfd2.phys.uni.lodz.pl}
\begin{abstract}
The $\gamma$/hadron separation in the imaging air Cherenkov telescope technique is based on differences between images of a hadronic shower and a $\gamma$ induced electromagnetic cascade. 
One may expect for a large telescope that a detection of hadronic events containing Cherenkov light from one $\gamma$ subcascade only is possible. In fact, simulations show that for the MAGIC telescope their fraction in the total protonic background is about 1.5$\%$ to 5.2$\%$ depending on the trigger threshold. It has been found that such images have small sizes (mainly below 400 photoelectrons) which correspond to the low energy primary $\gamma$'s (below 100 GeV). 
It is shown that parameters describing shapes of images from one subcascade have similar distributions to primary $\gamma$ events, so those parameters are not efficient in all methods of $\gamma$ selection. 
Similar studies based on MC simulations are presented also for the images from 2 $\gamma$ subcascades which are products of the same $\pi^0$ decay. 
The ratio of the number of the expected background from false $\gamma$ and one $\pi^0$ to the number of the triggered high energy photons from the Crab direction has been estimated for images with a small alpha parameter to show that the occurrence of this type of protonic shower is the reason for the difficulties with true $\gamma$ selection at low energies.
\end{abstract}

\submitto{\JPG}
\maketitle

\section{Introduction}

The first TeV $\gamma$-ray source, the Crab Nebula, was discovered by the Wipple collaboration in 1989 \cite{whipple} and began the very fast development of ground-based $\gamma$-ray astronomy. It was done by using a 10 m-diameter telescope which collected Cherenkov light produced in the atmosphere by charged fast particles of a shower produced by $\gamma$-ray. The light image of the shower was recorded by a matrix of photomultipliers mounted in the focal plane of the telescope (camera). In 1985 Hillas \cite{hillas} proposed a method of $\gamma$ selection from the hadronic background, several orders of magnitude larger. Parameters proposed by Hillas described the shape and the orientation of the image. The direction of the image main axis determines the direction of the primary particle; a narrow shape indicates a $\gamma$-ray as a primary particle.\\  
Imaging air Cherenkov telescopes (IACT) have been used for years to discover new $\gamma$ ray sources and determine their fluxes.
Larger telescopes have been built, stereoscopy is used to lower the energy thresholds and improve the sensitivity, but the main idea of $\gamma$/hadron separation remains the same. 
Major contributions to the development of the ground-based $\gamma$-ray astronomy have been made by recent experiments with very large reflectors like CANGAROO \cite{mori}, HESS \cite{hofmann}, MAGIC \cite{mariotti}, VERITAS \cite{weeks}.\\On the one hand very large telescopes make detection of low energy events possible, on the other hand they require a parabolic shape of the main reflector dish (like that in MAGIC) to avoid broadening the time profile of Cherenkov signal. Unfortunately these kind of telescopes, in spite of the Cotton-Devis design \cite{cotton}, have non-negligible aberrations - one of the sources of worsening the $\gamma$ selection. 
The $\gamma$/hadron separation becomes more difficult in the low energy region because of higher relative fluctuations in the shower development, causing larger fluctuations of the Cherenkov light density \cite{bhat} and image parameters.\\
As expected, telescopes may be triggered by light produced by one charge particle only, like a muon. When the image contains photons from one muon only and the impact parameter is larger than 80 m, then it is very similar to a $\gamma$ cascade image \cite{mirzoyan}. 
Rejecting such events is possible by using fast FADC in the readout system because they have a very narrow arrival time distribution \cite{mirzoyan}.\\
It can happen that all photons in a hadron-initiated shower registered by the telescope are produced by one electromagnetic subcascade. I shall call it a 'false $\gamma$ image'. 
The only physical reason for such images to be different from primary $\gamma$-induced events is that an average they may begin deeper in the atmosphere and in the result they may have slightly narrower angular distribution of charge particles and image shapes. 
Subcascades are created in the showers mostly by the $\pi^0$ decay process. It is also possible that an image contains photons from only two $\gamma$ subcascades, which are products of the same $\pi^0$ decay. I shall call it 'one $\pi^0$ image'. In this case images can be slightly wider but still might imitate primary $\gamma$-rays. However the main axis orientation of a false $\gamma$ or a one $\pi^0$ image should be random in contrast to true $\gamma$ events where the main axis is directed towards the centre of the camera (if the telescope is directed towards a $\gamma$-ray source).\\
In this article the fraction of the false $\gamma$ and one $\pi^0$ events in the proton showers is calculated by Monte Carlo simulations for the MAGIC telescope. 
In the following I present a MC study, discuss the similarity of the false and true $\gamma$ image parameters and draw conclusions about the worsening ability for $\gamma$ separation for low size images due to the occurrence of this type of background.\\

\section{ Experiment}
The MAGIC telescope \cite{albert} started operation at the end of 2003. It is situated in the Canary Island of La Palma (28.8N, 17.9W) at the Roque de los Muchachos Observatory, 2200 m above sea level. 
Currently it is the largest working IACT in the world. The telescope frame made from carbon fibre tubes, has a parabolical shape with a focal length of 17 m. 
The diameter of the telescope is 17 m. The main dish is covered by more than 900, 0.5 x 0.5 $m^2$  aluminium mirrors to achieve the total telescope area of 236 $m^2$. The inner part of the camera consists of 397 photomultipliers (ET 9116) with 0.1$^o$ diameter.
The additional 180 PMT (ET9117) with 0.2$^o$ diameter are the outer part of the camera. The hexagonal shape camera covers in total 4$^o$. There are 325 pixels (in the inner part) used for trigger. The trigger requires time coincidence and neighbour logic.

\section{Monte Carlo simulations}
The CORSIKA code version 6.023 \cite{heck,knapp} was used in simulations. GHEISHA and VENUS have been chosen as low (primary momentum below 80 GeV/c) and high energy interaction models. Proton-induced showers were simulated in the energy range 30 GeV - 1 TeV with the differential spectral index -2.75. The simulations were done for the telescope zenith angle of 20$^o$ and azimuth of 0$^o$ (showers directed to the north). Four different simulation sets have been performed with different cone opening angles of shower directions distributed isotropically inside the cone and impact parameter range. The overview of the number of simulated events and chosen limits in all MC sets are presented in Table 1.\\

\begin{table}
\caption {\label{tab1}Number of simulated proton showers and limits of cone opening angle and impact parameter.}
\begin{indented}
\item[]\begin{tabular}{@{}*{5}{l}}
\br
MC set & I & II & III & IV\cr
\mr
Inner limiting angle of view cone (deg.) & 0. & 0. & 3. & 1.0\cr
Outer limiting angle of view cone (deg.) & 3. & 3. & 5. & 5.5\cr
Inner limiting impact parameter (km) & 0.0 & 0.5 & 0. & 0.8\cr
Outer limiting impact parameter (km) & 0.5 & 0.8 & 0.8 & 1.0\cr
Number of simulated events in $10^6$ & 5.0 & 2.4 & 3.6 & 3.6\cr
\br
\end{tabular}
\end{indented}
\end{table}

In $\gamma$ cascade simulations the impact parameter was distributed uniformly in a circle with radius 300 m. The direction of 35000 simulated $\gamma$ events was fixed to a zenith angle of 20$^o$ and azimuth angle of 0$^o$ (parallel to the telescope axis) and the energy range was from 10 GeV to 30 TeV with a spectral index of -2.6, which is the index of the Crab spectrum (for energy above 100 GeV).\\
The standard CORSIKA code was modified to keep additional information about each subcascade produced in the EAS. The type of charge particle responsible for each Cherenkov photon creation is also known in this code.\\
The simulations were done for MAGIC as an example of a stand-alone air Cherenkov telescope \cite{albert,bario}. The Rayleigh and Mie scattering were taken into account according to the Sokolsky formula \cite{sokol}. The full geometry of the mirrors and their imperfections were considered in a reflector program. The next step of the detector simulations included the camera properties, such as additional reflections in Winston cones, photocathode quantum efficiency \cite{bario} and its fluctuation, and the photoelectron collection efficiency.\\
The night sky background (NSB) of $1.75 *10^{12}~ph/(m^2~sr~s)$, which was measured on La Palma \cite{nsb}, was included in the simulations before checking trigger conditions. However, photoelectrons produced by NSB have not been added to the image. Adding NSB to the images requires a cleaning procedure and by doing so at too high a level one may make images artificially narrower and hence the image parameters comparison less reliable.

To consider the simulated shower as a triggered event it was required that the output signals in the 4 next neighbouring pixels (4 NN) exceed thresholds in time gate of 3 ns. The proper discriminator simulations including arrival time of the photoelectrons and superimposed photoelectron pulses were done \cite{sob}. 
As trigger thresholds four possibilities were chosen: 4, 6, 8, 10 arbitrary units (from now I call it a.u.), which corresponds respectively to 4, 6, 8, 10 photoelectrons (p.e.) arriving at the same time. 

\section{Results and discussion}

Figure 1 shows the dependence of the angular distance between the telescope and shower axes (and in the figures it is call OFF) on the simulated impact parameter of proton events for the trigger threshold 4 a.u.. The smooth distribution in Figure 1a, which corresponds to all triggered events, confirms proper adding of all MC sets. 
Proton-induced showers whose images have no light contributions from hadrons or muons (I call them electromagnetic images or events), contain false $\gamma$ and one $\pi^0$ candidates and their distribution is shown in Figure 1b. The electromagnetic images occur more seldom for impact smaller than 100 m (first 2 bins of the impact parameter) because Cherenkov photons produced by hadrons and muons are expected for these distances from the core axis. 
Figures 1c and 1d present the same dependence for one $\pi^0$ and false $\gamma$. One $\pi^0$ events have an axis core position mostly below 800 m while false $\gamma$ images may appear as showers also with larger impact parameters.\\ 
\begin{figure}
\begin{center}
\includegraphics*[width=14cm]{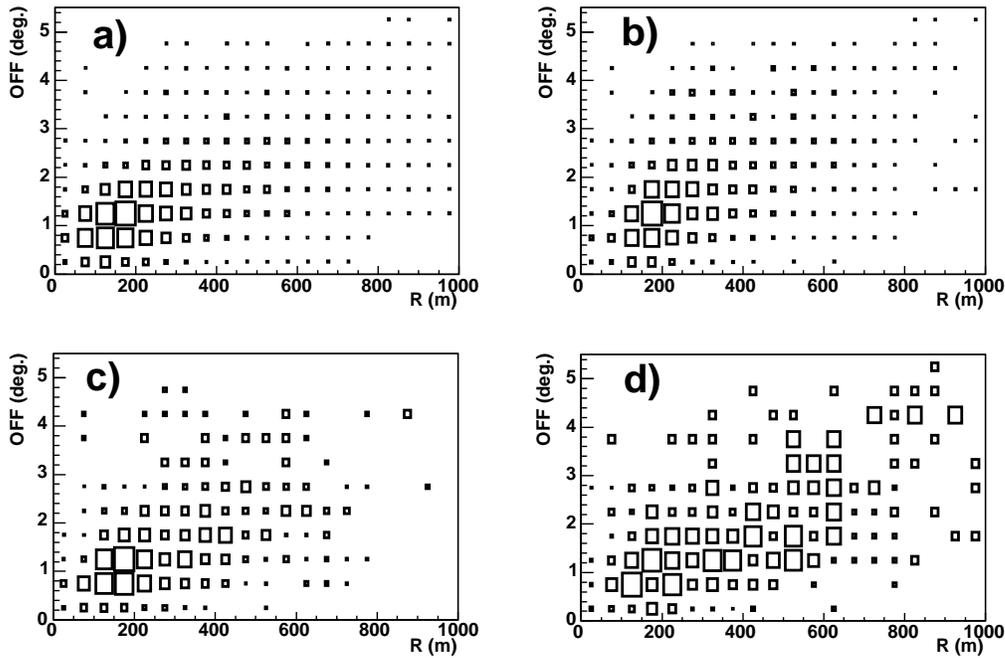}
\end{center}
\caption{Distribution of the angle between the shower and telescope axes and the impact parameter for trigger threshold 4 a.u.: {\bf a)} all triggered proton events; {\bf b)} electromagnetic; {\bf c)} one $\pi^0$; {\bf d)} false $\gamma$. The area of a square is $\sim$ to the number of events, but absolute numbers in each figure are different.}

\label{impact}
\end{figure}

The primary energy distribution of the simulated and triggered proton events, for trigger threshold 4 a.u., is shown in Figure 2. Five histograms in the figure correspond to the five cases: all simulated showers, all triggered events, the electromagnetic ones, one $\pi^0$ and false $\gamma$ events. In the case of showers, which may imitate true $\gamma$ (solid and dashed dotted lines in Figure 2), distributions are much steeper than in the case of all triggered events for primary energies above 100 GeV. Around 94$\%$ of all false $\gamma$'s have primary energy below 200 GeV. Primary energy of the proton is lower than 300 GeV for 96$\%$ of all one $\pi^0$ events.\\

\begin{figure}
\begin{center}
\includegraphics*[width=9cm]{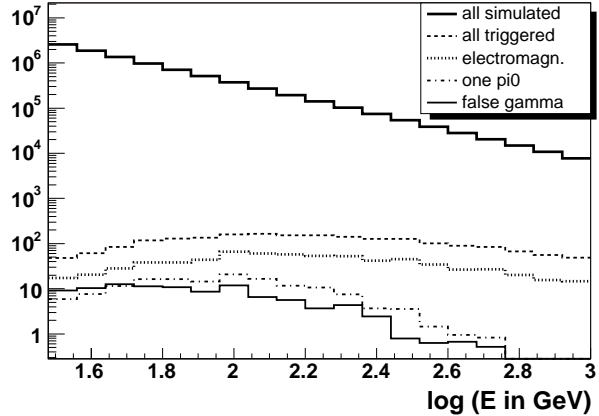}
\end{center}
\caption{Energy distribution for proton events (threshold 4 a.u.).}
\label{energy}
\end{figure} 

The most interesting question is how often the protonic CR background can produce false $\gamma$ and one $\pi^0$ images. This fraction may be estimated from the simulations because the probability of the triggering event with an impact parameter larger than 1000 m and declined more than 5.5$^o$ to the telescope axis is negligibly small in comparison to all triggered showers. The expected number of triggered events with primary energy above 1 TeV (maximum of the simulated energy) was estimated by extrapolation of the energy distribution plot to the energy 10 TeV. The simply power law fit of the distribution tail (above 600 GeV) was used as an extrapolation function. The same procedure was performed to estimate the expected number of false $\gamma$ and $\pi^0$ images. In this case the energy range from 100 GeV to 1 TeV has been used as the distribution tail. 
The dependence of the fraction of the false $\gamma$ and one $\pi^0$ on the trigger threshold is shown in Figure 3. 
The contribution of false $\gamma$ in all triggered events decreases from 5.2$\%$ to 1.5$\%$ with the increasing trigger threshold from 4 a.u. to 10 a.u.. The estimated fraction of one $\pi^0$ images changes from 6.6$\%$ to 3.2$\%$.\\
A similar fraction was calculated for electromagnetic events (and also their subclasses one $\pi^0$ and false $\gamma$), whose images contain 10$\%$ of light from muons or hadrons. The fraction of electromagnetic events in all triggered events is around 5$\%$ higher than in Figure 3 for all trigger thresholds. The rate of proton showers, which still may imitate true gamma's (both false $\gamma$ and one $\pi^0$), is about 0.3$\%$ higher than in Figure 3 for the lowest trigger threshold. For the highest trigger threshold this fraction does not change. The additional false $\gamma$ and one $\pi^0$ events have impact parameters below 400 m mostly and relatively small OFF (below 2.5$^o$). Hadrons and muons have negligible contribution to the calculated fraction for larger distances and OFF angles.\\

\begin{figure}
\begin{center}
\includegraphics*[width=9cm]{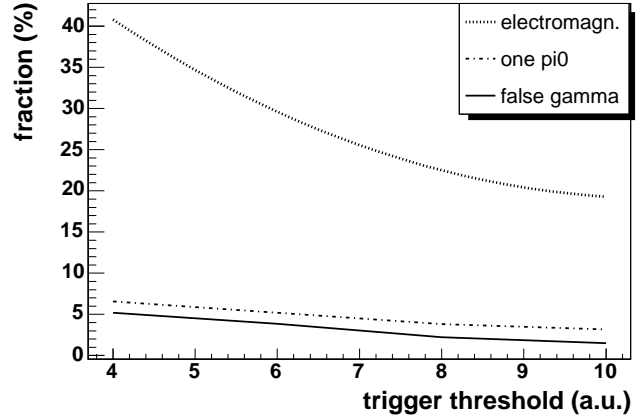}
\end{center}
\caption{Fraction of the interesting events in the total protonic background as a function of the trigger threshold.}
\label{fraction}
\end{figure} 

As described in the Monte Carlo section there is no influence of NSB to the image itself and no cleaning procedure was applied before calculating Hillas parameters in the present analysis. 
Figure 4 shows the size distribution for trigger threshold 4 a.u.. The image size is the sum of detected photoelectrons. Both false gammas and one $\pi^0$ images have small sizes because their primary energy is low. Apart from that around half of one subcascade events have a large impact parameter (above 400 m; see Figure 1c and 1d) where the expected size is low.\\

\begin{figure}
\begin{center}
\includegraphics*[width=9cm]{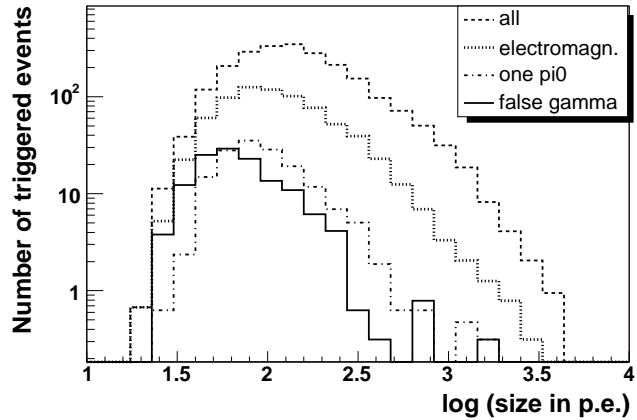}
\end{center}
\caption{Size distributions for proton showers; trigger threshold 4 a.u..}
\label{size}
\end{figure} 

It is seen in Figure 4 that the fraction of false $\gamma$ and one $\pi^0$ events in the triggered proton showers depends on the chosen size interval. One may estimate the energy of primary gammas, whose sizes are similar to the sizes of the background from false $\gamma$ and one $\pi^0$ events. As such we chose the peak position in the energy distribution of the primary gammas for the given size interval. The results are summarized in Table 2 for trigger threshold 4 a.u.. The expected background from both false $\gamma$ and one $\pi^0$ events is 4.5$\%$ for energy of true $\gamma$'s around 100 GeV and it increases significantly at lower energies. 
We may expect that even more than 25$\%$ of all triggered proton showers (with image size below 100 p.e) are detected as false $\gamma$ or one $\pi^0$ events.\\

\begin{table}
\caption {\label{tab2}The fraction of events unrecognised by shape in the total sample of the triggered proton showers for five size bins (trigger threshold 4 a.u.). The last column corresponds to the position of the energy peak for $\gamma$ simulations.}

\begin{indented}
\item[]\begin{tabular}{@{}*{7}{llllll}}
\br                              
Size &false $\gamma$&one $\pi^0$ & the energy peak\\
(in p.e.)& (in $\%$)&( in $\%$) & (in GeV)\\
\mr
$<$100&12.7&12.8& 30\\
(100,160)&4.1&7.6& 50\\
(160,250)&2.5&5.0& 70\\
(250,400)&1.0&3.5& 100\\
$>$400&0.5&2.5.& 160\\
\br
\end{tabular}
\end{indented}
\end{table} 
The next image parameter which is important for this analysis is the width. 
The width is one of the Hillas parameters describing the image shape.
The comparison of the width distributions of the simulated $\gamma$, false $\gamma$ and one $\pi^0$ images is presented in Figure 5a. No cleaning procedure was applied to MC simulations to show the effect of shower development only. The distribution of one $\pi^0$ events looks a little shifted (towards larger widths) because the angular distribution of charged particles in one $\pi^0$ events is wider than in one $\gamma$ cascade. The directions of two gammas after $\pi^0$ decay are not the same. The separation angle depends on the energy of $\pi^0$ and the way of the energy sharing to the products of decay. For example a $\pi^0$ with energy higher than 20 GeV decays to two gammas with separation angle comparable to the opening angle of the Cherenkov light cone.\\ 
There is an excess of events in the distribution of false $\gamma$'s for low widths caused by a detection of subcascdes which begin deeper in the atmosphere. In this case the observed cascade parts are younger so that the angular distribiution of electrons and positrons is narrower than in the primary gamma cascades (see e.g. \cite{giller}), resulting in lower widths.\\ 
Different OFF distributions are also the reason for differences between the two width distributions. Figure 5b shows the comparison of width distributions of the true gamma's and both false $\gamma$'s and one $\pi^0$'s, which have an OFF angle below 1$^o$. For small OFF angles primary $\gamma$ and false $\gamma$ or $\pi^0$ do not differ in width distribution, so this parameter is not a good variable to select true $\gamma$'s from the part of the proton background which is discussed in this paper.\\

\begin{figure}
\begin{center}
\includegraphics*[width=14cm]{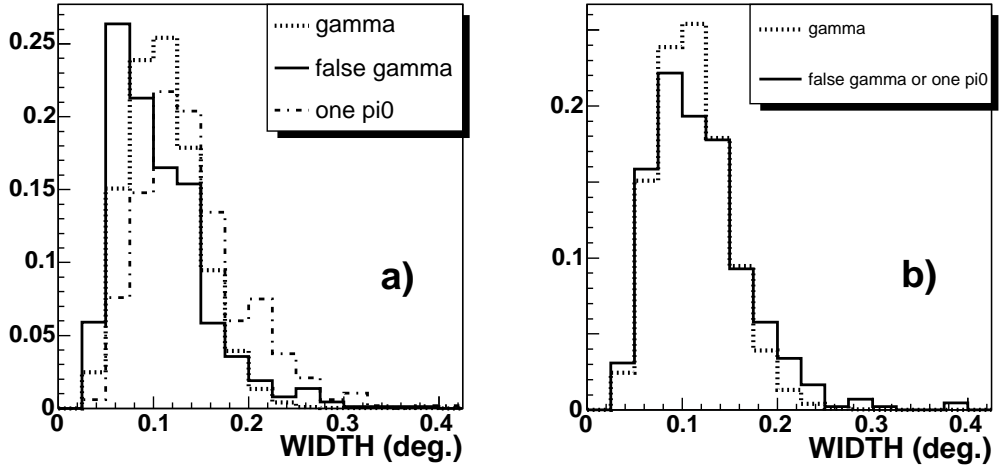}
\end{center}
\caption{{\bf a)} Width distributions for the true $\gamma$, false $\gamma$ and one $\pi^0$ events, for trigger threshold 4 a.u.; {\bf b)} The same as in a) but with OFF angle below 1$^o$. All histograms are normalized to 1.}
\label{width}
\end{figure} 

Another Hillas parameter describing the image shape is length. The length distributions of images corresponding to true $\gamma$, false $\gamma$ and one $\pi^0$ are shown in Figure 6a. In the case of true $\gamma$'s the distribution is narrower than for false $\gamma$ and one $\pi^0$. This fact can be explained by different inclination angles. The directions of the simulated primary $\gamma$'s are parallel to the telescope axis and only a part of the longitudinal cascade is detected by the telescope. False $\gamma$ and one $\pi^0$ are inclined to the telescope axis and it may happen that a longer part of the shower is visible, resulting in a larger image length. 
The opposite scenario is also possible: the observed longitudinal part of a shower is shorter than in case of a true $\gamma$ and then false $\gamma$ image has a smaller length. The second effect is not visible in one $\pi^0$ events because the products of $\pi^0$ decay are separated by a non negligible angle.\\
Figure 6b shows the comparison of length distributions for primary $\gamma$'s and both the false $\gamma$'s and one $\pi^0$'s with a small OFF angle (below 1$^o$). The differences seen in Figure 6a have almost disappeared. Thus, the reason for them are different arrival direction distributions of proton showers and primary $\gamma$ cascades.\\

\begin{figure}
\begin{center}
\includegraphics*[width=14cm]{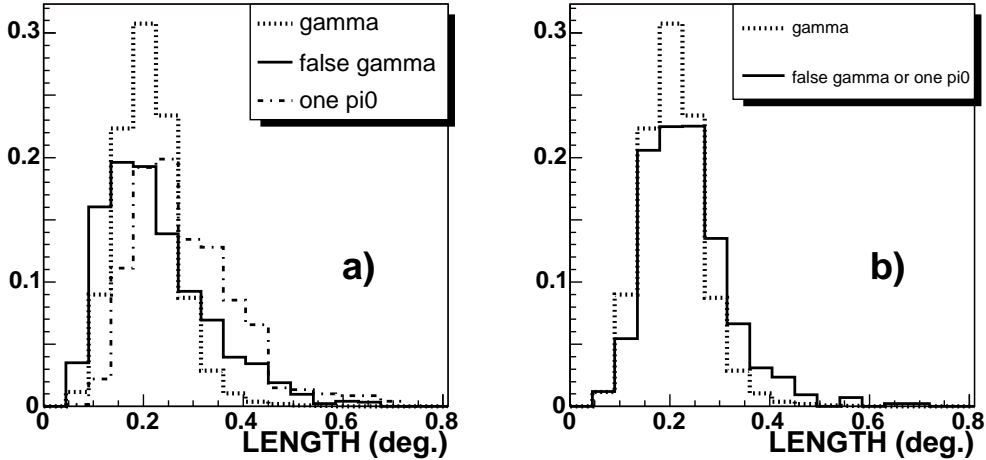}
\end{center}
\caption{{\bf a)} Length distributions for true $\gamma$, false $\gamma$ and one $\pi^0$ events, for trigger threshold 4 a.u.; {\bf b)} The same as in a) but with OFF angle below 1$^o$. All histograms are normalized to 1.}

\label{length}
\end{figure} 

There is a Hillas parameter called alpha, which is connected with the primary particle direction. Different OFF distributions of proton and $\gamma$ events are the reason for the expected differences of the alpha parameter distributions.
We have checked that the distributions of the alpha parameter are uniform for all cases: electromagnetic, one $\pi^0$ or false $\gamma$. This parameter can be still efficiently used in $\gamma$/hadron separation. However the final selection (using alpha parameter) depends on the chosen size interval because images imitating $\gamma$-rays exist and their fraction is size dependent. The ratio of the expected number of both false $\gamma$ and one $\pi^0$ images to that of primary $\gamma$'s from the Crab Nebula direction has been calculated in different size bins. The only alpha cut (below 15$^o$) has been applied to these calculations. The background was estimated from simulations using primary proton and He spectra  \cite{bess,hareyama}. To add primary He (not simulated) I have assumed that they produce the same fraction of the false $\gamma$ and one $\pi^0$ as primary protons. The expected number of true $\gamma$'s from Crab was estimated from the simulations using the spectrum measured by MAGIC \cite{wagner} and extrapolating it to lower energies. The Crab spectrum below 100 GeV is probably flatter but the power law extrapolation seems to be good enough for this estimation because the real spectrum is unknown in this energy region. The ratio of the expected number of false $\gamma$ and one $\pi^0$ images to the expected number of the true $\gamma$'s (from Crab) for trigger threshold 4, 6 and 8 a.u. is shown in Figure 7. This ratio depends on the hardware trigger thresholds for low size events, but it is almost stable and remains on the level 1 for sizes above 400 p.e.. For all trigger threshold cases the lower is the size the higher is the number of false $\gamma$ and one $\pi^0$ images (with alpha parameter below 15$^o$) detected in the same time as one high energy photon from Crab (with alpha parameter below 15$^o$). In case of trigger threshold 4 a.u. the detection of one true gamma with size below 100 p.e. is associated with detection of about nine false $\gamma$ or one $\pi^0$ events with sizes below 100 p.e.. The occurrence of this type of the hadronic events results in the substantial limitation of the $\gamma$/hadron separation for images in the low size range.

\begin{figure}
\begin{center}
\includegraphics*[width=9cm]{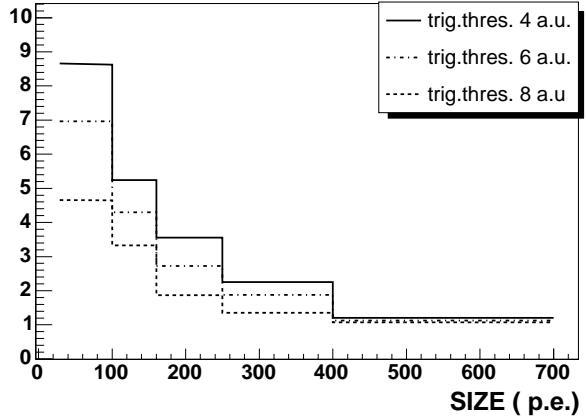}
\end{center}
\caption{Ratio of the expected number of false $\gamma$ and one $\pi^0$ images to the expected number of the true $\gamma$ (from Crab) in different size bins for alpha parameter below 15$^o$.}
\label{fralpha}
\end{figure}

\section {Conclusion}
Our simulations show that images produced by Cherenkov photons from one electromagnetic subcascade in proton-induced showers occur in large telescopes. 94$\%$ of such events correspond to the primary energy of protons lower than 200 GeV in the case of the MAGIC telescope for trigger thresholds from 4 a.u.. Their impact parameter may be very large and there is no special dependence on the inclination angle of the shower axis to that of the telescope. 
The fraction of such events in the total proton background decreases with increasing trigger threshold from 5.2$\%$ to 1.5$\%$ for thresholds from 4 to 10 a.u.. It was shown that sizes of such events are small, mostly below 400 photoelectrons. The images of these events have shapes very similar to those from primary $\gamma$'s.\\
It was also shown that images from two subcascades from the same $\pi^0$ appear in the trigger. Their fraction is estimated from 6.6$\%$ to 3.2$\%$ for trigger thresholds from 4 to 10 a.u.. The images are a little wider and longer than those of true $\gamma$'s. Efficient true $\gamma$ separation from one $\pi^0$ and false $\gamma$ images is not possible by using the shape parameters only.\\ 
It is suggested in \cite{maier} that this hardly reducible background is caused by hadronic showers in which most of the primary energy is transfered to electromagnetic subcascades during the first few interactions. They are mainly diffractive interactions, so that the variables estimated in this paper depend on the interaction model. The GHEISHA model which gives the lowest probability that more of the 50$\%$ primary energy is deposited in the electromagnetic part of the shower \cite{maier} has been used in the presented analysis. Our simulations using even such a model show that more than 25$\%$ of protonic images with sizes below 100 p.e. are detected as one subcascade or one $\pi^0$ events (see Table 2).
Thus, in the low energy region, below $\sim$ 100 GeV, $\gamma$/hadron separation is much more difficult due to the existence of the false $\gamma$ and one $\pi^0$ images. We may expect this background to survive in the alpha distribution even after selection procedures based on image shapes. In the experiment the eventual background at small alpha parameters is estimated by an extrapolation of the background at larger alpha. However the telescope sensitivity deteriorates significantly at low energies due to the presented natural limit on the $\gamma$-ray selection.

\ack
This work was supported by Polish KBN grant No. 1P03D01028.

\end{document}